\documentclass[a4paper]{jpconf}
\usepackage{subeqn} 
\usepackage[square,sort&compress]{natbib}

\newcommand{\bt}{{\bf t}}

\newcommand{\bgamma}{\boldmath{\gamma}}

\newcommand{\bS}{{\bf S}}
\newcommand{\bT}{{\bf T}}

\begin{document}
\title{The tensor of interaction of a two-level system with an arbitrary strain field}
\author{D V Anghel$^{1,2}$, T K\"uhn$^3$, Y. M. Galperin$^{4,5,6}$ and M. Manninen$^3$}
\address{$^1$ Department of Theoretical Physics, National Institute for Physics and Nuclear Engineering--''Horia Hulubei'', Bucharest - Magurele, Romania\\
$^2$ Bogoliubov Laboratory of Theoretical Physics, JINR Dubna, Russia\\
$^3$ Nanoscience Center, Department of Physics, University of Jyv\"askyl\"a,  Jyv\"askyl\"a, Finland\\
$^4$ Department of Physics \& Centre of Advanced Materials and Nanotechnology,
University of Oslo, Oslo, Norway\\
$^5$ Argonne National Laboratory, 9700 S. Cass Av., Argonne, IL 60439, USA\\
$^6$ A. F. Ioffe Physico-Technical Institute of Russian Academy of Sciences, 194021 St. Petersburg, Russia}

\begin{abstract}
The interaction between two-level systems (TLS) and strain fields in a solid 
is contained in the diagonal matrix element of the interaction hamiltonian, 
$\delta$, which, in general, has the expression $\delta=2[\gamma]:[S]$, 
with the tensor $[\gamma]$ describing the TLS ``deformability'' and 
$[S]$ being the symmetric strain tensor. We construct $[\gamma]$ on very 
general grounds, by associating to the TLS two objects: a direction, 
$\hat\bt$, and a forth rank tensor of coupling constants, $[[R]]$. 
Based on the method of construction and on the invariance of the 
expression of $\delta$ with respect to the symmetry 
transformation of the solid, we conclude that $[[R]]$ has the same structure 
as the tensor of stiffness constants, $[[c]]$, from elasticity theory. 
In particular, if the solid is isotropic, $[[R]]$ has only two independent 
parameters, which are the equivalent of the Lam\'e constants. Employing this 
model we calculate the absorption and emission rates of phonons on TLSs 
and show that in isotropic solids, on average, the longitudinal phonons 
interact stronger with the TLSs than the transversal ones, as it is 
observed in experiments. We also show that in isotropic solids, a transversal 
wave leaves unperturbed all the TLSs with the direction contained in one of 
the two planes that are 
perpendicular either to the wave propagation direction or to the polarization 
direction and that a longitudinal strain applied to the solid polarises 
the TLS ensemble. 
\end{abstract}

In a temperature range around ten Kelvins or below, the physical 
properties of amorphous materials differ significantly from the 
properties of crystals and show striking universal features 
(see \cite{esquinazi:book} for a collection of reviews). Most of these 
features can be explained by assuming that in the 
amorphous solid exists a collection of dynamic defects, which are 
atoms or groups of atoms, oscillating in two well potentials. 
At low temperatures the thermal activation is suppressed and the 
oscillation happens by quantum tunnelling 
from one potential minimum to the other, forming in this way what is 
called a \textit{two-level system} (TLS). In the
two-dimensional Hilbert space  spanned by the ground states of the 
two wells, the effective Hamiltonian of the TLS is
\begin{equation}
  \label{eqn_TLS_hamiltonian}
  H_{TLS}
      =  \frac{\Delta}{2}\sigma_z -\frac{\Lambda}{2}\sigma_x 
 \equiv \frac{1}{2}\left(\begin{array}{cc} 
                     \Delta   & -\Lambda\\ 
                     -\Lambda & -\Delta \end{array}
                \right) 
\end{equation}
where $\sigma_x$ and $\sigma_z$ are Pauli matrices, while $\Delta$ and 
$\Lambda$ are called the {\em asymmetry of the potential} 
and the {\em tunnel splitting}, respectively. The Hamiltonian 
(\ref{eqn_TLS_hamiltonian}) may be diagonalized by an orthogonal 
transformation $O$, 
%
%
$H'_{TLS}\equiv O^T H_{TLS} O = \frac{\epsilon}{2}\sigma_z$, to obtain 
the excitation energy of the TLS, 
$\epsilon\equiv\sqrt{\Delta^2+\Lambda^2}$--by the superscript $T$ we denote 
in general \textit{the transpose of a matrix}. 
The parameters $\Delta$ and $\Lambda$ do not have the same values 
for all the TLSs, but are distributed 
with the density $VP(\Delta,\Lambda)$, where $V$ is the volume of the solid. 
According to the standard tunneling model (STM), this distribution is 
$P(\Delta,\Lambda)=P_0/\Lambda$, 
%
%
with $P_0$ a constant.
If expressed through the variables $\epsilon$ and $u\equiv\Lambda/\epsilon$,
the distribution function becomes 
%
%
$P(\epsilon,u)= P_0/(u\sqrt{1-u^2})$. 

A phonon, or any other strain in the solid body, perturbes 
$H_{TLS}$ by $H_1\equiv(\delta/2)\sigma_z$ \cite{esquinazi:book,JLowTempPhys.7.351.Philips,PhilMag.25.1.Anderson,ZPhys.257.212.1972.Jackle,RevModPhys.59.1.Leggett}:
%
%
The perturbation $\delta$ is linear in the strain field, $S_{ij}$, 
\cite{RevModPhys.59.1.Leggett,esquinazi:book} and in general may be written 
as $\delta\equiv 2\gamma_{ij} S_{ij}$--here, as everywhere in this paper, 
we assume \textit{summation over repeated indices}.
The $3\times 3$ symmetric strain tensor is defined as 
$S_{ij}=\frac{1}{2}(\partial_iu_j+\partial_ju_i)$, with $u_i$ ($i=1,2,3$) 
being the components of the displacement field. In dyadic notations, 
$\delta=2[\gamma]:[S]$, where by $[\cdot]$ we shall denote matrices or 
second rank tensors. 

In the construction of $[\gamma]$ we have to be more careful. In the STM,
if $H_1$ is used to describe the interaction of TLS with phonons 
in three-dimensional (3D) bulk systems, 
the tensor $[\gamma]$ is replaced by a scalar, $\gamma_l$ or
$\gamma_t$, depending whether the phonon has longitudinal or transversal 
polarization, respectively, while $S$ takes the value of the amplitude 
of the strain field. So one would write $\delta=2(\gamma_lS_l+\gamma_tS_t)$. 

This picture cannot be applied to the interaction of TLS with arbitrary 
strain fields, at least because simple coordinate transformations may 
lead to ambiguity. It is well known from elasticity theory that by 
coordinate transformations longitudinal stress can be transformed into 
shear stress and vice-versa. To avoid this ambiguity, we have to keep 
$\gamma$ in the form of a tensor, not a simple matrix, and try to find its 
components \cite{PhysRevB75.064202.2007.Anghel}. 
So, the question we 
ask ourselves is how can we extract, on very general grounds, 
the tensor of interaction constants from the physical characteristics 
of the TLS--note that a tensor changes its components at the 
coordinate transformations while the value of $\delta$ should be the same 
in any coordinate system. Where can we begin to construct such a 
tensor? 

Following the line of arguments from \cite{PhysRevB75.064202.2007.Anghel}, 
we start by noting that since the TLS is represented as a 
tunnelling entity between two potential wells--the transition from 
one classical equilibrium position to the other taking place either 
by a rotation or by a translation--we may associate to the TLS a 
direction in space, $\hat\bt$. So we 
have now three components, $t_1$, $t_2$, $t_3$, which change under 
coordinates transformations. The simplest $3\times3$ symmetric tensor 
that we can construct 
from $\hat\bt$ is $[T]=\hat\bt\cdot\hat\bt^T$ ($T_{ij}=t_it_j$), while 
a general one would have the components $\gamma_{kl}=R_{ijkl}T_{ij}$. 
In abbreviated subscript notations (see for example \cite{auld:book} 
for details), $\bT=(t_x^2,t_y^2,t_x^2,2t_yt_z,2t_zt_x,2t_xt_y)^T$, 
and $R_{ijkl}$ becomes $R_{IJ}$ in a straightforward way; in these 
notations $[\gamma]$ becomes the vector $\bgamma\equiv[R]^T\cdot\bT$. 
Since $[S]$ is also transformed into 
$\bS\equiv(S_{xx},S_{yy},S_{zz},2S_{yz},2S_{zx},2S_{xy})^T$, the component 
of the interaction hamiltonian, $\delta$, is written simply as 
$\delta\equiv2\bT^T\cdot[R]\cdot\bS$, which, we say it again, is a scalar 
under coordinates transformations. 

Now notice the analogy between $\delta$ and the elastic energy density, $u$, 
that exists in a deformed body. In abbreviated subscript notations, 
$u=\frac{1}{2}\bS^T\cdot[c]\cdot\bS$, where $[c]$ is the $6\times6$ 
matrix of the elastic stiffness constants. Under a coordinate transformation, 
$\bS$ transforms into $\bS'=[N]\cdot\bS$, where $[N]$ is the $6\times6$ 
matrix of the transformation (see Ref. \cite{auld:book}, Eq. 3.34), so 
\begin{equation}
u=u'=\frac{1}{2}(\bS')^T\cdot[c]\cdot\bS'
=\frac{1}{2}\bS^T\cdot[N]^T\cdot[c]\cdot[N]\cdot\bS.
\label{utransf}
\end{equation}
Since (\ref{utransf}) should be valid for any $\bS$ and any transformation, 
then $[c]$ should remain 
unchanged--$[c]=[N]^T\cdot[c]\cdot[N]$--under the symmetry 
transformations of the crystalline lattice. From this argument follow 
all the properties of the matrix $[c]$ which are characteristic to 
the symmetries of the crystal under consideration \cite{auld:book}. 

The same is true for $\delta$. Here $\bT$ transforms in the same way as 
$\bS$ under coordinates transformations--$\bT'=[N]\cdot\bT$. 
Therefore $\delta\equiv\delta'=\bT^T\cdot[N]^T\cdot[R]\cdot[N]\cdot\bS$ and, 
like for $[c]$, we have $[R]=[N]^T\cdot[R]\cdot[N]$ for any symmetry 
transformation $[N]$. 
In conclusion, $[R]$ has the same structure as the tensor of elastic stiffness 
constants (see also \cite{PhysRevB75.064202.2007.Anghel}).

If a material is isotropic, $[c]$ has only two independent 
parameters--the \textit{Lam\'e constants}, $\lambda$ and $\mu$: 
$c_{IJ}=2\mu\delta_{IJ}+\lambda$ for $I,J\le 3$  and 
$c_{IJ}=\delta_{IJ}\mu$ for $I$ or $J$ bigger than 3. In normal 
subscripts $c_{ijkl}$ can be written in the more compact form, $c_{ijkl}=\lambda\delta_{ij}\delta_{kl}+\mu(\delta_{ik}\delta_{jl}+\delta_{il}\delta_{jk})$. 
Based on the arguments above, the same is true for $[R]$; let us 
denote the independent parameters of $[R]$ by $\zeta$ and $\xi$, and, in 
normal subscripts, $R_{ijkl}=\zeta\delta_{ij}\delta_{kl}+\xi(\delta_{ik}\delta_{jl}+\delta_{il}\delta_{jk})$. 
For a more straightforward reference to the STM, it might be even more 
convenient to denote $\tilde\gamma\equiv\zeta+2\xi$ and introduce the 
reduced tensor $[[r]]$ by $r_{ijkl}\equiv R_{ijkl}/\tilde\gamma=\zeta'\delta_{ij}\delta_{kl}+\xi'(\delta_{ik}\delta_{jl}+\delta_{il}\delta_{jk})$, 
where $\zeta'+2\xi'=1$. 

Using such a form of the coupling constants tensor, we calculated in 
\cite{PhysRevB75.064202.2007.Anghel} the scattering rate 
of phonons on the TLSs and showed that we recover the results of the STM 
after we average over the directions of the TLS, assuming that they 
are isotropically oriented. The two constants of the STM, $\gamma_l$ 
and $\gamma_t$, are related to the parameters in this model by 
$\gamma_l=\tilde\gamma C_l$ and $\gamma_t=\tilde\gamma C_t$, with 
\begin{equation}
C_l=\frac{1}{15}(15-40\xi'+32(\xi')^2)\qquad {\rm and}\qquad 
C_t=\frac{4}{15}(\xi')^2. \label{ClandCt}
\end{equation}
From Eqs. (\ref{ClandCt}) we see immediately that $C_l>C_t\ge0$ for any real 
$\xi$, as it is observed experimentally in general 
\cite{PhysRevB75.064202.2007.Anghel}. 

What is also interesting to note, is that by calculating 
$\gamma_l$ and $\gamma_t$ from the experimental data, we can calculate 
$\xi$ and $\zeta$ and determine completely the tensor of coupling 
constants, $[R]$. 
For example the two different sets of values for $P_0\gamma_l$ and 
$P_0\gamma_l$ in fused silica,
reported by Golding et al \cite{PhysRevB.14.1660.Golding} 
($P_0\gamma_l^2=1.4\times10^{-5}$ J/m$^3$ and 
$P_0\gamma_t^2=0.63\times 10^{-5}$ J/m$^3$) and 
Hunklinger and Arnold \cite{PhysicalAcousticsHunklinger} 
($P_0\gamma_l^2=2.0\times 10^{-5}$
J/m$^3$ and $P_0\gamma_t^2=0.89\times 10^{-5}$ J/m$^3$), 
cited also by Black \cite{PhysRevB.17.2740.1978.Black},
give the same solutions for $\xi'$: $\xi'_1=0.55$ and $\xi'_2=1.2$. 

If we now go back to the interaction of a single TLS with a strain 
field and we assume that a transversal wave of strain field 
$\bS=(0,0,0,S,0,0)^T$, is propagating through the solid, we 
find $\delta=4\tilde\gamma\xi't_yt_zS$. This means that any TLS 
of $t_y=0$ or $t_z=0$ (i.e. any TLS which is contained in a plane 
perpendicular either to the propagation direction or to the 
polarization direction) will not be perturbed by this wave. 

Observing that for both values of $\xi'$ calculated 
above, the corresponding values of $\zeta'(=1-2\xi')$ are negative, we show 
an interesting polarization effect of the TLS ensemble, due to an external 
stress. Let us assume that we apply a longitudinal stress along the 
$z$ direction, $\bS=(0,0,S,0,0,0)^T$. This stress gives a perturbation 
$\delta = 2\tilde\gamma t_z^2St_z^2+2\tilde\gamma\zeta'S(t_x^2+t_y^2)$ and we 
notice that the two terms in the expression of 
$\delta$--$\delta_1=2\tilde\gamma t_z^2St_z^2$ and 
$\delta_2=22\tilde\gamma\zeta'(t_x^2+t_y^2)S$--have opposite signs. This 
means that, if e.g. $\tilde\gamma t_z^2S>0$, the energy splitting, and 
therefore the excitation energy, of the TLSs oriented along the strain 
increase, while the energy splitting and the excitation energy of the 
TLSs oriented perpendicular to the strain direction decrease. In other 
words, the strain polarises the TLS ensemble. 

In conclusion, we used a model for the interaction of two-level 
systems (TLS) with arbitrary strain fields, introduced in 
\cite{PhysRevB75.064202.2007.Anghel}, which assumes that 
to any TLS is associated a direction, $\hat\bt$, and a tensor of coupling 
constants to the strain field, $[[R]]$, and we showed on general grounds 
that $[[R]]$ has the same structure with respect to the symmetry 
transformations of the solid as the tensor of stiffness constants, $[[c]]$, 
from the elasticity theory. Some immediate consequences of this formalism 
are that in isotropic solids, on average, the longitudinal phonons 
interact stronger with the TLSs than the transversal ones, 
($\gamma_l>\gamma_t$, in the language of the standard tunnelling model), 
a transversal wave does not interact with the TLSs contained in one of 
the two planes that are 
perpendicular either to the wave propagation direction or to the polarization 
direction, and a strain applied to the body may polarize 
the TLS ensemble. 

\ack

This work was partly supported by the U. S. Department of Energy Office of
Science under the Contract No. DE-AC02-06CH11357 and by the NATO grant 
EAP.RIG 982080. DVA acknowledges the hospitality of the University of 
Jyv\"askyl\"a, where part of this work has been done, and 
the financial support from the Academy of Finland. 

\providecommand{\newblock}{}

\end{document}